\documentclass[onecolumn,showpacs,amsmath,amssymb,12pt]{revtex4}
\usepackage{graphicx}
\usepackage{dcolumn}
\usepackage{bm}
\begin{document}
%\preprint{APS/123-QED}
%%%%%%%%%%%%%%%%%%%%%%%%
\newcommand{\hs}{\hspace*{0.5cm}}
\newcommand{\vs}{\vspace*{0.5cm}}
\newcommand{\be}{\begin{equation}}
\newcommand{\ee}{\end{equation}}
\newcommand{\bea}{\begin{eqnarray}}
\newcommand{\eea}{\end{eqnarray}}
\newcommand{\ben}{\begin{enumerate}}
\newcommand{\een}{\end{enumerate}}
\newcommand{\bde}{\begin{widetext}}
\newcommand{\ede}{\end{widetext}}
\newcommand{\nn}{\nonumber}
\newcommand{\crn}{\nonumber \\}
\newcommand{\Tr}{\mathrm{Tr}}
\newcommand{\non}{\nonumber}
\newcommand{\noi}{\noindent}
\newcommand{\al}{\alpha}
\newcommand{\la}{\lambda}
\newcommand{\bet}{\beta}
\newcommand{\ga}{\gamma}
\newcommand{\va}{\varphi}
\newcommand{\om}{\omega}
\newcommand{\pa}{\partial}
\newcommand{\+}{\dagger}
\newcommand{\fr}{\frac}
\newcommand{\bc}{\begin{center}}
\newcommand{\ec}{\end{center}}
\newcommand{\Ga}{\Gamma}
\newcommand{\de}{\delta}
\newcommand{\De}{\Delta}
\newcommand{\ep}{\epsilon}
\newcommand{\varep}{\varepsilon}
\newcommand{\ka}{\kappa}
\newcommand{\La}{\Lambda}
\newcommand{\si}{\sigma}
\newcommand{\Si}{\Sigma}
\newcommand{\ta}{\tau}
\newcommand{\up}{\upsilon}
\newcommand{\Up}{\Upsilon}
\newcommand{\ze}{\zeta}
\newcommand{\ps}{\psi}
\newcommand{\Ps}{\Psi}
\newcommand{\ph}{\phi}
\newcommand{\vph}{\varphi}
\newcommand{\Ph}{\Phi}
\newcommand{\Om}{\Omega}
%%%%%%%%%%%%%%%%%%%%%%%%

\title{The 3-3-1 model with inert scalar triplet}

\author{P. V. Dong}
\email {pvdong@iop.vast.ac.vn} \affiliation{Institute of Physics, Vietnam Academy of Science and Technology, 10 Dao Tan, Ba Dinh, Hanoi, Vietnam}

\author{T. Phong Nguyen}
\email{thanhphong@ctu.edu.vn}
\affiliation{Department of Physics, Cantho University, 3/2 Street, Ninh Kieu, Cantho, Vietnam}

\author{D. V. Soa}
\email{dvsoa@assoc.iop.vast.ac.vn}
\affiliation{Department of Physics, Hanoi National University of Education, 136 Xuan Thuy, Cau Giay, Hanoi, Vietnam}

\date{\today}

\begin{abstract}
We show that the typical 3-3-1 models are only self-consistent if they contain interactions explicitly violating the lepton number. The 3-3-1 model with right-handed neutrinos can by itself work as an economical 3-3-1 model as a natural recognition of the above criteria while it also results an inert scalar triplet ($\eta$) responsible for dark matter. This is ensured by a $Z_2$ symmetry (assigned so that only $\eta$ is odd while all other multiplets which perform the economical 3-3-1 model are even), which is not broken by the vacuum. The minimal 3-3-1 model can provide a dark matter by a similar realization. Taking the former into account, we show that the dark matter candidate ($H_\eta$) contained in $\eta$ transforms as a singlet in effective limit under the standard model symmetry and being naturally heavy. The $H_\eta$ relic density and direct detection cross-section will get right values when the $H_\eta$ mass is in TeV range as expected. The model predicts the $H_\eta$ mass $m_{H_\eta}=\la_5\times 2$ TeV and the $H_\eta$-nucleon scattering cross-section $\sigma_{H_\eta-N}=1.56\times 10^{-44}\ \mathrm{cm}^2$, provided that the new neutral Higgs boson is heavy enough than the dark matter.           
\end{abstract}

\pacs{12.60.-i, 95.35.+d}

\maketitle

\section{\label{intro}Introduction}       

The standard model has been very successful in describing the world of fundamental particles and interactions among them \cite{pdg}. Notably, the Higgs particle, a long-standing hypothesized scalar that consequently provides the masses for all other particles, has finally been proved by the recent CERN-LHC experiments, where the new discovered resonance is standard model like \cite{atlas,cms}. However, the standard model fails to answer a large portion of the total mass-energy of the universe such as dark matter $>$ 20\% and dark energy $>$ 70\%, which all lie beyond the standard model particle content \cite{pdg}. 

The well-motivated theories that by themselves result dark matter as a consequence can be listed as supersymmetry \cite{martin}, extradimension \cite{extrareview} or little Higgs model \cite{littlereview}. In a recent article, we added to the list by showing that the dark matter can also naturally be resulted from the 3-3-1-1 gauge theory by itself \cite{dongdm} (a theory that originally provides the potential explanations of fermion generation number \cite{anoma}, uncharacteristically-heavy top quark \cite{longvan}, strong $CP$ \cite{palp}, and electric charge quantization \cite{ecq}). Indeed, this 3-3-1-1 gauge symmetry that includes $B-L$ (baryon minus lepton number) as its residual and non-commuting gauge charge is the necessary extension of 3-3-1 models~\cite{331m,331r,dongfla} that respect the conservation of lepton and baryon numbers, similarly to the case of electric charge operator. In other words, this new theory of strong, electroweak and $B-L$ interactions is a direct consequence of non-closed algebra between $B-L$ and 3-3-1 symmetry~\cite{dongdm}. Consequently, the conserved and unbroken $W$ parity, similarly to $R$ parity in supersymmetry, can be resulted as a residual symmetry of broken 3-3-1-1 gauge symmetry or more detailed $B-L$ (this breaking possibly happens at a scale matching the 3-3-1 breaking scale of TeV order that makes the model consistent without the necessity of a large desert as in grand unified theories \cite{su5,so10}). Among the existing 3-3-1 models, we have found that the most new particles of the 3-3-1 model with neutral fermions~\cite{dongfla}, the so-called $W$ particles, transform nontrivially (that is odd) under the $W$ parity, which are responsible for dark matter~\cite{dongdm}. 

By contrast, all the new particles in the 3-3-1 model with right-handed neutrinos~\cite{331r} as well as those of the minimal 3-3-1 model \cite{331m} are even under the $W$ parity. Therefore, the $W$ parity transforms trivially that is useless for these models in responsibility to the problem of dark matter \cite{dongdm}. On the other hand, it is well-known that the 3-3-1 model with right-handed neutrinos might actually accommodate potential candidates for dark matter~\cite{dm331}.  However, all the extra symmetries studied therein (which had existed before the $W$ parity) such as the $Z_2$, lepton charge, or even a generic continuous symmetry if imposed for their stability are subsequently violated or broken, which leads to the fast decay of dark matter, as explicitly shown in~\cite{dongdm} (this will be also extensively analyzed below before concluding this work). Hence, it is necessary to find a new mechanism rather than the useless $W$ parity and the mentioned extra symmetries that is responsible for the dark matter stability in the 3-3-1 model with right-handed neutrinos. This mechanism should also be applicable to the minimal 3-3-1 model for a realization of dark matter (notice that this model has previously been predicted containing no dark matter, by contrast). 

To proceed further, we first suppose that the lepton number in the 3-3-1 model with right-handed neutrinos is an approximate symmetry, which avoids the gauged symmetry of the lepton number nor the 3-3-1-1 extension \cite{dongdm}. This proposal indeed realizes a theory that explicitly violates the $W$ parity or lepton number symmetry in order to make it (our 3-3-1 model) self-consistent. Exactly, the 3-3-1 model with right-handed neutrinos often works with three scalar triplets $\rho$, $\eta$ and $\chi$, where $\eta$ and $\chi$ transform identically under the 3-3-1 gauge symmetry. However, the $\eta$ and $\chi$ differ in the lepton charge~\cite{lepto331}. Since the lepton charge symmetry is already violated, these two scalars can act as equivalent representations under any mentioned group that operate on the model. We could therefore remove one of them from the theory (assumed $\eta$). The result is a new, consistent model working with only two scalar triplets $\rho$, $\chi$ which explicitly recognizes the violation of $W$ parity or lepton number symmetry. This theory has been extensively studied over the last decade and named as the economical 3-3-1 model \cite{ecn331}. However, the economical 3-3-1 model does not contain any dark matter too, which is unlike the conclusion of \cite{hltk}.      

In this work, by contrast to that approach \cite{ecn331} we will retain the $\eta$ in the theory, but study how it is hidden instead of removing it. For this aim, we first assume that the $\eta$ transforms odd under a $Z_2$ symmetry, whereas $\chi$, $\rho$ and all other fields are even (notice that this $Z_2$ differs from the one mentioned above). We then prove that the vacuum can be stabilized, conserving the $Z_2$ symmetry. The lightest particle resided in the ``inert'' scalar triplet $\eta$ is thus stabilized responsible for dark matter, while the remaining scalars $\rho$, $\chi$ develop the vacuum expectation values (VEVs) for breaking the gauge symmetry and mass generation in a correct way like the economical 3-3-1 model. This approach is completely distinguished from the previous studies \cite{dm331,dongdm} because it is based on the economical 3-3-1 model (with lepton number violation responsible for neutrino masses) as a consequence other than the 3-3-1 model with right-handed neutrinos (with the lepton number conserved which is unrealistic). Also, its results---the dark matter candidate and phenomenology as recognized are entirely different from that of the inert doublet model \cite{idmma} as well as those in \cite{dm331,dongdm}. By the same way, the minimal 3-3-1 model can behave as a reduced 3-3-1 model \cite{rm331} while containing an inert scalar triplet responsible for dark matter.      
   
The rest of this paper is organized as follows. In Sec. \ref{model} we propose the new model. We first give a discussion on lepton number, its violation, and introduce the $Z_2$ symmetry and inert scalar triplet. We then consider the gauge symmetry breaking and prove that the $Z_2$ is unbroken by the vacuum. The candidates of dark matter which lie in the scalar sector are identified, and their interactions are obtained. Section \ref{relicdensity} is devoted to the dark matter relic density and dark matter constraints due to direct searches. Section \ref{why} is plausible to point out why our work is necessary, newly achieved and its implication to other 3-3-1 models. Finally, we summarize our results and make conclusions in the last section---Sec. \ref{conclusion}.

\section{\label{model}The model}

\subsection{Lepton number violation, $Z_2$ symmetry and inert scalar triplet}

The model under consideration is based on the $SU(3)_C\otimes SU(3)_L\otimes U(1)_X$ (3-3-1) gauge symmetry. The fermion content is given by \cite{331r}  
\bea \psi_{aL} &\equiv & \left(\begin{array}{c}
               \nu_{aL}\\ e_{aL}\\ (\nu_{aR})^c
\end{array}\right) \sim (1,3, -1/3),\hs e_{aR}\sim (1,1, -1),
\\
Q_{1L} &\equiv& \left(\begin{array}{c} u_{1L}\\  d_{1L}\\ U_L \end{array}\right)\sim
 \left(3,3,1/3\right),\hs Q_{\al L}\equiv \left(\begin{array}{c}
  d_{\al L}\\  -u_{\al L}\\  D_{\al L}
\end{array}\right)\sim (3,3^*,0), \\ u_{a
R}&\sim&\left(3,1,2/3\right),\hs d_{a R} \sim
\left(3,1,-1/3\right),\\ U_{R}&\sim& \left(3,1,2/3\right),\hs D_{\al R}
\sim \left(3,1,-1/3\right),\eea  where the quantum numbers defined in the parentheses are given upon the $(SU(3)_C$, $SU(3)_L$, $U(1)_X)$ symmetries, respectively. The family indices are set as $a=1,2,3$ and $\al=2,3$. The $\nu_{aR}$ are the right-handed neutrinos which are correspondingly included to complete the lepton triplet representations (thus the model named 3-3-1 model with right-handed neutrinos). Similarly, the exotic quarks $U$, $D_\al$ take in part of the respective quark multiplets. The last two families of quarks that transform under $SU(3)_L$ differently from the first one and those of leptons are arranged in order to cancel the $SU(3)_L$ self-anomaly (i.e. the number of fermion triplets must be equal to that of antitriplets). It is easily checked that the theory is free from all the other anomalies. 

The electric charge operator, which is only the generator conserved after the gauge symmetry breaking, is given by \be Q=T_3-\fr{1}{\sqrt{3}}T_8+X,\ee where $T_i\ (i=1,2,...,8)$ is the charge of $SU(3)_L$, while $X$ is that of $U(1)_X$ (below, the $SU(3)_C$ charges will be denoted by $t_i$). Let us note that the exotic quarks $U$ and $D_\al$ have electric charges like ordinary quarks, $Q(U)=2/3$ and $Q(D_\al)=-1/3$, respectively. 

The baryon number ($B$) as a global symmetry $U(1)_B$ commutes with the gauge symmetry and being always conserved by the general Lagrangian and vacuum \cite{lepto331}. However, the lepton number~($L$) of lepton triplet components is given by $(+1,+1,-1)$ which does not commute with the gauge symmetry, similarly to the case of electric charge. In addition, the algebra of $L$ and 3-3-1 symmetry is non-closed because in order for $L$ to be some generator of $SU(3)_L\otimes U(1)_X$, i.e. $L=x_i T_i + y X$ with fixed $x_i$, $y$ coefficients, we have $\mathrm{Tr}(L) = y \mathrm{Tr}(X)$ for every multiplet, which is generally incorrect \cite{dongdm}. For examples, we have $y=-1$ for $e_R$, but $y=0$ for $u_R$, which contradicts. Therefore, if the lepton number $L$ is conserved, we can find in the resulting theory an extra $U(1)_{\mathcal{L}}$ group factor so that its Lagrangian is invariant under this group, and the combination obtained \cite{lepto331} \be L=\fr{4}{\sqrt{3}}T_8+\mathcal{L},\label{leptocharge} \ee as a residual charge of $SU(3)_L\otimes U(1)_{\mathcal{L}}$. The $\mathcal{L}$-charges for fermion multiplets are given by \bea \mathcal{L}(\psi_{aL})&=&1/3,\ \mathcal{L}(Q_{1L})=-2/3,\ \mathcal{L}(Q_{\al L})=2/3, \crn \mathcal{L}(e_{aR})&=&1,\ \mathcal{L}(u_{aR})=\mathcal{L}(d_{aR})=0,\ \mathcal{L}(U_R)=-2,\ \mathcal{L}(D_{\al R})=2.\eea In addition, the exotic quarks satisfy $L(U)=-L(D)=-2$, which are called leptoquarks. 

Notice that the above definition (\ref{leptocharge}) is only given if the lepton number of the theory is conserved. Otherwise, the Yukawa Lagrangian and scalar potential will take the most general forms which have interactions explicitly violating $L$ \cite{lepto331}. Also, the $L$ is subsequently broken. There is no $U(1)_{\mathcal{L}}$ at all. The relation (\ref{leptocharge}) disappears, which avoids the judgment of \cite{dongdm} (there, the 3-3-1 model that conserves $L$ must be extended to the 3-3-1-1 model with gauged $B-L$ as a result of gauged $T_8$). This is a new observation of this work which is to be studied below (in other words, the 3-3-1 model is only self-consistent by this case of lepton-number violation). Namely, the lepton number will not be regarded as an exact symmetry of the theory; however, we can consider it as an approximate symmetry to keep the model self-consistent. Therefore, the (\ref{leptocharge}) is only approximate expression for calculating the lepton number of model particles (because the theory is obviously not constrained to be invariant under the approximate symmetry $U(1)_{\mathcal{L}}$, as supposed). It is noteworthy that this charge is thus no longer to be regarded as a gauge symmetry as in~\cite{dongdm}. All the above ingredients can also be applied to the minimal 3-3-1 model. A theory that does not satisfy the criteria of \cite{dongdm} is the economical 3-3-1 model \cite{ecn331}. In the present work we are going to realize a new 3-3-1 model of this kind.     

As usual, the 3-3-1 model with right-handed neutrinos requires three scalar triplets \cite{331r},
\bea
\eta &=&  \left(\begin{array}{c}
\eta^0_1\\
\eta^-_2\\
\eta^0_3\end{array}\right)\sim (1,3,-1/3),\label{vev1}\\
\phi &=& \left(\begin{array}{c}
\phi^+_1\\
\phi^0_2\\
\phi^+_3\end{array}\right)\sim (1,3,2/3),\label{vev2}\\ 
 \chi &=& \left(\begin{array}{c}
\chi^0_1\\
\chi^-_2\\
\chi^0_3\end{array}\right)\sim (1,3,-1/3),\label{vev3}\eea to break the gauge symmetry and generating the masses. Hereafter, we use the notation $\phi$ instead of $\rho$ mentioned in the introduction so that it is similar to that of the economical 3-3-1 model. The $\mathcal{L}$-charges for the scalar triplets are obtained by \cite{lepto331} \be \mathcal{L}(\chi)=4/3,\ \mathcal{L}(\phi)=-2/3,\ \mathcal{L}(\eta)=-2/3.\ee The nonzero lepton numbers of scalars are \be L(\chi^0_1)=L(\chi^-_2)=-L(\phi^+_3)=-L(\eta^0_3)=2.\ee Because the lepton number is an approximate symmetry, all the electrically-neutral scalars including bileptons $\chi^0_1$ and $\eta^0_3$ might develop VEVs as given in the next subsection. The electroweak gauge symmetry is broken via two stages. In the first stage, $SU(3)_L \otimes U(1)_X$ is broken down to that of the standard model generating the masses of new particles. This is achieved by the VEV of $\chi^0_3$ (possibly that of $\eta^0_3$ too). In the second stage, the standard model electroweak symmetry is broken down to $U(1)_Q$ responsible for the masses of ordinary particles. This stage is achieved by the VEV of $\phi^0_2$ and/or $\eta^0_1$ (possibly that of $\chi^0_1$ too).  

Let us remind the reader that the $\eta,\ \chi$ have the same gauge quantum numbers. They differs only in the $\mathcal{L}$-charge as given above. Since the $U(1)_{\mathcal{L}}$ symmetry is approximate just as its violating interactions allowed, they equivalently act on the model. This results the economical 3-3-1 model that works with only two scalar triplets $\phi,\ \chi$ by excluding the $\eta$ \cite{ecn331}. In this article, we will introduce another scenario that we retain the $\eta$ in the theory but impose a $Z_2$ symmetry so that the only $\eta$ is odd: 
\be \eta \rightarrow -\eta.\ee All the other multiplets including $\phi,\ \chi$ are even, $\phi\rightarrow \phi,\ \chi \rightarrow \chi$, and so on. 

Up to the gauge fixing and ghost terms, the Lagrangian is given by
\bea \mathcal{L}&=&\sum_{\mathrm{Fermion\ multiplets}}\bar{F}i\ga^\mu D_\mu F + \sum_{\mathrm{Scalar\ multiplets}}(D^\mu S)^\dagger (D_\mu S)\crn
&& -\fr{1}{4}G_{i\mu\nu}G_i^{\mu\nu} -\fr 1 4 A_{i\mu\nu}A_i^{\mu\nu}-\fr 1 4 B_{\mu\nu} B^{\mu\nu} + \mathcal{L}_{\mathrm{Y}} - V,\label{latoanphan}\eea with the covariant derivative $D_\mu = \pa_\mu + ig_s t_i G_{i\mu} + ig T_i A_{i\mu} + i g_X (X/\sqrt{6}) B_\mu$, and the field strength tensors $G_{i\mu\nu}=\pa_\mu G_{i\nu} -\pa_\nu G_{i\mu} -g_s f_{ijk}G_{j\mu}G_{k\nu}$, 
$A_{i\mu\nu}=\pa_\mu A_{i\nu} -\pa_\nu A_{i\mu} - g f_{ijk}G_{j\mu} A_{k\nu}$,
and $B_{\mu\nu}=\pa_\mu B_\nu -\pa_\nu B_\mu$, correspondingly to the $SU(3)_C$, $SU(3)_L$ and $U(1)_X$ groups. The last two terms will be specified below.  

The Yukawa Lagrangian is given by 
\bea \mathcal{L}_{\mathrm{Y}}&=&h^e_{ab}\bar{\psi}_{aL}\phi e_{bR}+ h^\nu_{ab}\bar{\psi}^c_{a L}\psi_{b L} \phi \crn
&& + h^U\bar{Q}_{1L}\chi U_R + h^D_{\al \beta}\bar{Q}_{\al L} \chi^* D_{\beta R}+h^d_a\bar{Q}_{1L}\phi d_{aR} +h^u_{\al a } \bar{Q}_{\al L}\phi^* u_{aR} \crn
&& + \bar{h}^u_a \bar{Q}_{1L}\chi u_{aR}+ \bar{h}^d_{\al a} \bar{Q}_{\al L}\chi^* d_{aR}+\bar{h}^D_\al \bar{Q}_{1L}\phi D_{\al R} +\bar{h}^U_{\al } \bar{Q}_{\al L}\phi^* U_{R} +H.c. \eea 
Due to the $Z_2$ symmetry, the $\eta$ does not interact with fermions. The Yukawa Lagrangian is achieved similarly to the economical 3-3-1 model \cite{ecn331}. The couplings $\bar{h}$s violate the lepton number, while the $h$s do not. The fermions get consistent masses at the one-loop level, or alternatively via the five-dimensional effective interactions \cite{ecn331}. Below, we will prove $\langle \eta \rangle =0$. Therefore, the gauge bosons get masses from the vacuum values of $\phi$ and $\chi$, which are similar to the economical 3-3-1 model too.    

The scalar potential that is invariant under the gauge symmetry, the $Z_2$, and being renormalizable is given by 
\bea V&=&\mu^2_1\phi^\dagger \phi +\mu^2_2 \chi^\dagger \chi + \mu^2_3 \eta^\dagger \eta\crn
&&+\la_1(\phi^\dagger \phi)^2 +\la_2 (\chi^\dagger \chi)^2 + \la_3 (\eta^\dagger \eta)^2\crn
&&+\la_4(\phi^\dagger \phi)(\chi^\dagger \chi)+\la_5(\phi^\dagger \phi)(\eta^\dagger \eta)+\la_6(\chi^\dagger \chi)(\eta^\dagger \eta)\crn
&&+\la_7(\phi^\dagger \chi)(\chi^\dagger \phi)+\la_8(\phi^\dagger \eta)(\eta^\dagger \phi)+\la_9(\chi^\dagger \eta)(\eta^\dagger \chi)\crn
&&+\fr 1 2 [\la_{10}(\eta^\dagger \chi)^2+H.c.] \eea Here, $\mu^2_{1,2,3}$ and $\la_{1,2,3,...,9}$ are real whereas $\la_{10}$ can be complex. However, the phase of $\la_{10}$ can be removed by redefining the relative phases of $\eta$ and $\chi$. Consequently, this potential conserves $CP$ symmetry. But the $CP$ symmetry can be broken spontaneously by the VEVs of the scalars. It is also noted that the coupling $\la_{10}$ violates the lepton number \cite{lepto331}. 

If we point out that the minimization of the above scalar potential conserves the $Z_2$ symmetry, i.e. $\langle \eta \rangle =0$, the $Z_2$ is exact and unbroken. Consequently, the $\eta$ is only coupled in pairs, in interacting with the economical 3-3-1 model particles. This proposal already realizes a 3-3-1 model with ``inert'' scalar triplet ($\eta$). The lightest particle contained in $\eta$ is absolutely stabilized which can be responsible for dark matter. The inert particles are naturally recognized by the original scalar sector of the 3-3-1 model with right-handed neutrinos \cite{331r}. By contrast, in the inert doublet model \cite{idmma} the similar one should be introduced to the standard model by hand. It is easily realized that the $\phi$ and $\eta$ contain two scalar doublets---the one in $\phi$ is similar to the standard model doublet, while another in $\eta$ is the inert doublet. However, it is noted that due to the gauge symmetry the $\eta$ is not coupled to $\phi$ via a coupling similarly to $\la_{10}$, which is unlike the inert doublet model. Hence, the dark matter phenomenology in our theory is completely distinguished as shown below.           

\subsection{Gauge symmetry breaking and $Z_2$ conservation}

Since the lepton number is violated, all the neutral scalars can develop VEVs. Assume that the scalar potential is minimized at 
\bea \langle \phi \rangle = (0,v_\phi,0),\hs \langle \chi \rangle = (u_\chi,0,\om_\chi),\hs \langle \eta \rangle = (u_\eta,0,\om_\eta),\eea with its value 
\bea V_{\mathrm{min}} &=& \mu_1^2 v^*_\phi v_\phi +\mu^2_2 (u^*_\chi u_\chi+\om^*_\chi \om_\chi)+\mu_3^2 (u^*_\eta u_\eta+\om^*_\eta \om_\eta)\crn 
&&+\la_1(v^*_\phi v_\phi)^2+\la_2(u^*_\chi u_\chi+\om^*_\chi \om_\chi)^2+\la_3 (u^*_\eta u_\eta+\om^*_\eta \om_\eta)^2\crn
&&+\la_4(v^*_\phi v_\phi)(u^*_\chi u_\chi+\om^*_\chi \om_\chi)+\la_5 (v^*_\phi v_\phi)(u^*_\eta u_\eta+\om^*_\eta \om_\eta)\crn
&&+\la_6 (u^*_\eta u_\eta+\om^*_\eta \om_\eta)(u^*_\chi u_\chi+\om^*_\chi \om_\chi)\crn 
&&+ \la_9 (u^*_\eta u_\chi+\om^*_\eta \om_\chi)(u^*_\chi u_\eta+\om^*_\chi \om_\eta)\crn
&& +\fr 1 2 [\la_{10}(u^*_\eta u_\chi+ \om^*_\eta \om_\chi)^2+H.c.]\eea The conditions of potential minimization are therefore given by 
\bea \fr{\pa V_{\mathrm{min}}}{\pa v^*_\phi}&=&v_\phi [\mu^2_1+2\la_1 (v^*_\phi v_\phi)+\la_4 (u^*_\chi u_\chi+\om^*_\chi \om_\chi)+\la_5 (u^*_\eta u_\eta+\om^*_\eta \om_\eta)]=0,\label{the}\\
\fr{\pa V_{\mathrm{min}}}{\pa u^*_\chi}&=&u_\chi [\mu^2_2+2\la_2(u^*_\chi u_\chi+\om^*_\chi \om_\chi)+\la_4(v^*_\phi v_\phi)+\la_6(u^*_\eta u_\eta+\om^*_\eta \om_\eta)]\crn
&&+u_\eta [\la_9(u^*_\eta u_\chi+\om^*_\eta \om_\chi)+\la^*_{10}(u^*_\chi u_\eta+\om^*_\chi \om_\eta)]=0,\label{the1}\\ 
\fr{\pa V_{\mathrm{min}}}{\pa \om^*_\chi}&=&\om_\chi [\mu^2_2+2\la_2(u^*_\chi u_\chi+\om^*_\chi \om_\chi)+\la_4(v^*_\phi v_\phi)+\la_6(u^*_\eta u_\eta+\om^*_\eta \om_\eta)]\crn
&&+\om_\eta [\la_9(u^*_\eta u_\chi+\om^*_\eta \om_\chi)+\la^*_{10}(u^*_\chi u_\eta+\om^*_\chi \om_\eta)]=0,\label{the2}\\
\fr{\pa V_{\mathrm{min}}}{\pa u^*_\eta}&=&u_\eta [\mu^2_3+2\la_3(u^*_\eta u_\eta+\om^*_\eta \om_\eta)+\la_5(v^*_\phi v_\phi)+\la_6(u^*_\chi u_\chi+\om^*_\chi \om_\chi)]\crn
&&+u_\chi [\la_9(u^*_\chi u_\eta+\om^*_\chi \om_\eta)+\la_{10}(u^*_\eta u_\chi+\om^*_\eta \om_\chi)]=0,\label{the3}\\
\fr{\pa V_{\mathrm{min}}}{\pa \om^*_\eta}&=&\om_\eta [\mu^2_3+2\la_3(u^*_\eta u_\eta+\om^*_\eta \om_\eta)+\la_5(v^*_\phi v_\phi)+\la_6(u^*_\chi u_\chi+\om^*_\chi \om_\chi)]\crn
&&+\om_\chi [\la_9(u^*_\chi u_\eta+\om^*_\chi \om_\eta)+\la_{10}(u^*_\eta u_\chi+\om^*_\eta \om_\chi)]=0.\label{the4}\eea 
Let us denote 
\bea A&=&\mu^2_2+2\la_2(u^*_\chi u_\chi+\om^*_\chi \om_\chi)+\la_4(v^*_\phi v_\phi)+\la_6(u^*_\eta u_\eta+\om^*_\eta \om_\eta),\crn
A'&=&\mu^2_3+2\la_3(u^*_\eta u_\eta+\om^*_\eta \om_\eta)+\la_5(v^*_\phi v_\phi)+\la_6(u^*_\chi u_\chi+\om^*_\chi \om_\chi),\crn
B&=&\la_9(u^*_\eta u_\chi+\om^*_\eta \om_\chi)+\la^*_{10}(u^*_\chi u_\eta+\om^*_\chi \om_\eta).\nn\eea The (\ref{the1}), (\ref{the2}), (\ref{the3}) and (\ref{the4}) are rewritten as 
\bea \left(\begin{array}{cc}
      u_\chi & u_\eta \\
      \om_\chi & \om_\eta \\
   \end{array}\right)
   \left(\begin{array}{c}
   A \\
   B\\
   \end{array}
   \right) &=& 0,\label{he1}\\
   \left(\begin{array}{cc}
      u_\chi & u_\eta \\
      \om_\chi & \om_\eta \\
   \end{array}\right)
   \left(\begin{array}{c}
   B^* \\
   A'\\
   \end{array}
   \right)&=&0.\label{he2}
\eea 

First of all, we suppose that the scalar potential is bounded from below. The necessary conditions are given by \be \la_1>0,\hs \la_2>0,\hs \la_3>0, \ee which can be obtained when $\phi$, $\chi$ or $\eta$ separately tending to infinity, respectively. To have a desired vacuum structure, we assume $\mu^2_{1,2} < 0$, $\mu^2_3 > 0$, $\la_{5}>0$, and $\la_6>0$. The last three conditions are given so that $A'>0$ (this will rearrange the general vacuum of the 3-3-1 model with right-handed neutrinos into the new one where its $Z_2$-even part is similar to the economical 3-3-1 model, while its $Z_2$-odd part conserves the $Z_2$ symmetry). Hence, from (\ref{he2}) we have $u_\eta/u_\chi=\om_\eta/\om_\chi \equiv t$. And, the (\ref{he1}) and (\ref{he2}) are reduced to
\bea A+tB=0,\hs B^*+t A'=0.\label{aba}\eea The second equation is rewritten by \be t\left[A'+\la_9\left(|u_\chi|^2+|\om_\chi|^2\right)\right]+t^*\la_{10}\left(|u_\chi|^2+|\om_\chi|^2\right)=0,\ee which implies $t=0$, thus $u_\eta=\om_\eta=0$, provided that $\la_6+\la_9\pm\la_{10}>0$ (to have such an unique solution). We have also $B=0$ and $A=0$ with the help of (\ref{aba}). Combined with the equation (\ref{the}), we have the solution of potential minimization as summarized below
\bea
|v_\phi|^2 &=& \fr{2\la_2 \mu^2_1-\la_4\mu^2_2}{\la^2_4-4\la_1\la_2} \neq 0,\label{dkvev1}\\ 
|u_\chi|^2 +|\om_\chi|^2&=& \fr{2\la_1\mu^2_2-\la_4 \mu^2_1}{\la^2_4-4\la_1\la_2} \neq 0,\label{dkvev2}\\
u_\eta &=&\om_\eta=0.\label{dkvev3} \eea We need extra conditions for the couplings,  
\be -2\sqrt{\la_1\la_2}<\la_4<\mathrm{Min}\left\{2\la_1\left({\mu_2}/{\mu_1}\right)^2,2\la_2\left({\mu_1}/{\mu_2}\right)^2\right\},\hs \la_7>0,\label{nhan1}\ee  These conditions have been achieved to make sure that the right-hand sides of (\ref{dkvev1}) and (\ref{dkvev2}) as well as the physical scalar masses given below are positive. They are also needed for the scalar potential bounded from below, when $\phi$ and $\chi$ simultaneously tend to infinity.   

It is easily realized that a part of the relations as given above is similar to the economical 3-3-1 model~\cite{ecn331}. Because of $\langle \eta\rangle =0$, the $Z_2$ symmetry is conserved by the vacuum also. Therefore, this symmetry is exact and being not spontaneously broken, similarly to $R$-parity in supersymmetry. Consequently, the so-called ``inert'' scalar triplet $\eta$, the only multiplet in the model that is charged under the $Z_2$ (odd), behaves similarly as superparticles in supersymmetry, which is distinguished from the remaining sector of $Z_2$-even normal matter. The lightest inert particle (LIP) contained in the $\eta$ triplet, which cannot decay due to the $Z_2$ symmetry, may provide dark matter candidates. On the other hand, as mentioned, the $\eta$ does not couple to fermions in the Yukawa sector, while it can couple to gauge bosons and other scalars via the $Z_2$ conserving interactions. It does not give masses for the fermions as well as gauge bosons since $\langle\eta\rangle=0$. The identification and mass of physical fermions and gauge bosons are exactly the same economical 3-3-1 model~\cite{ecn331}. However, the scalar sector will be changed that is presented below. The interaction between the two sectors, $Z_2$-even and odd, will be also obtained.  

For convenience in reading, let us redefine $u\equiv u_\chi$, $\om\equiv \om_\chi$, and $v\equiv v_\phi$. Thus, we have 
\bea \langle \phi \rangle = (0,v,0),\hs \langle \chi \rangle = (u,0,\om),\hs \langle \eta \rangle = (0,0,0),\label{dvacuumd}\eea where $v$, $u$ and $\om$ satisfy the potential minimization conditions (\ref{dkvev1}) and (\ref{dkvev2}), with the labels ``$\phi$'' and ``$\chi$'' removed. To keep a consistency with the standard model, we suppose $u^2\ll v^2\ll \om^2$, where $v=v_{\mathrm{weak}}=174\ \mathrm{GeV}$, $u=\mathcal{O}(1)\ \mathrm{GeV}$, and $\om=\mathcal{O}(1)\ \mathrm{TeV}$ \cite{ecn331}. Also, the conditions for the scalar potential parameters as obtained can be summarized as follows
\bea &&\mu^2_{1,2}<0<\mu^2_3,\hs \la_{1,2,3,5,6,7}>0,\hs \la_6+\la_{9}\pm\la_{10}>0, \crn 
&& -2\sqrt{\la_1\la_2}<\la_4<\mathrm{Min}\left\{2\la_1\left({\mu_2}/{\mu_1}\right)^2,2\la_2\left({\mu_1}/{\mu_2}\right)^2\right\}.\eea As mentioned, this ensures (i) the potential is bounded from below, (ii) the physical scalar masses are positive, (iii) the $Z_2$ symmetry is conserved by the vacuum, and (iv) the nonzero VEVs, $v$, $u$ and $\om$, induce the gauge symmetry breaking and mass generation in the correct way, similarly to the economical 3-3-1 model \cite{ecn331}.                  

\subsection{Scalar identification, dark matter and interactions}

The mass terms of physical scalar fields are obtained from the scalar potential by shifting the vacuum values of beginning scalars. They are given by 
\bea V_{\mathrm{mass}}&=& M^2_3\eta^\dagger \eta +\la_8 |v|^2\eta^-_2\eta^+_2+\la_9|u^*\eta^0_1+\om^*\eta^0_3|^2+\fr 1 2 [\la^*_{10}(u^*\eta^0_1+\om^*\eta^0_3)^2+H.c.]\crn
&&+\la_1(v^*\phi^0_2+v\phi^{0*}_2)^2+\la_2(u^*\chi^0_1+\om^*\chi^0_3+u\chi^{0*}_1+\om\chi^{0*}_3)^2\crn
&&+\la_4(v^*\phi^0_2+v\phi^{0*}_2)(u^*\chi^0_1+\om^*\chi^0_3+u\chi^{0*}_1+\om\chi^{0*}_3)\crn
&&+\la_7(v^*\chi^-_2+u\phi^-_1+\om \phi^-_3)(v\chi^+_2+u^*\phi^+_1+\om^* \phi^+_3),\eea
where $M^2_3\equiv \mu^2_3+\la_5|v|^2+\la_6(|u|^2+|\om|^2)$, and the conditions of potential minimization as given above have been used. Also, the notations of physical scalar fields have been taken the same outsets, which should be understood. 

{\it Inert scalar sector $(\eta)$:} The $h^\pm_\eta\equiv \eta^\pm_2$ is physical charged inert-scalar field by itself with the mass given by  
\be m^2_{h^\pm_{\eta}}=M^2_3+\la_8 v^2.\ee For the remaining inert fields let us define
\be \eta^0_1=\fr{R_1+i I_1}{\sqrt{2}},\hs \eta^0_3=\fr{R_3+i I_3}{\sqrt{2}}.\ee The mass Lagrangian for the neutral inert-scalar fields is arranged as \be \fr 1 2 (R_1\ I_1\ R_3\ I_3)M^2\left(\begin{array}{c}
R_1\\ I_1\\ R_3\\ I_3\end{array}\right),\ee in which the mass matrix $M^2$ is obtained by
\bea 
\left(\begin{array}{cccc}
M^2_3+\la_9 |u|^2 & \mathrm{Im}(\la_{10} u^2) & \mathrm{Re}[\om(\la_9 u^*+\la_{10}u)] & \mathrm{Im}[u(\la_{10}\om-\la_9\om^*)]\\ 
+\mathrm{Re}(\la_{10}u^2) & & & \\ [+4mm]
\mathrm{Im}(\la_{10} u^2) & M^2_3+\la_9|u|^2&\mathrm{Im}[u(\la_{10}\om+\la_9\om^*)]&\mathrm{Re}[\om(\la_9 u^*-\la_{10}u)]  \\ 
 & -\mathrm{Re}(\la_{10}u^2) & & \\ [+4mm]
\mathrm{Re}[\om(\la_9 u^*+\la_{10}u)] &\mathrm{Im}[u(\la_{10}\om+\la_9\om^*)]& M^2_3+\la_9 |\om|^2& \mathrm{Im}(\la_{10}\om^2)\\ 
& &+\mathrm{Re}(\la_{10}\om^2) & \\ [+4mm]
\mathrm{Im}[u(\la_{10}\om-\la_9\om^*)]&\mathrm{Re}[\om(\la_9 u^*-\la_{10}u)]&  \mathrm{Im}(\la_{10}\om^2)&M^2_3+\la_9 |\om|^2\\ 
& & & -\mathrm{Re}(\la_{10}\om^2)
\end{array}\right).\nn\eea It is recalled that the scalar potential conserves $CP$, so we can consider $\la_{10}$ to be real (otherwise its phase can be absorbed by redefining the relative phases of $\eta$ and $\chi$ as mentioned). In addition, the vacuum structure as obtained does not support any spontaneous $CP$ phase, i.e. the $CP$ symmetry is not spontaneously broken by the VEVs in this case. Therefore, without loss of generality we can assume that $u$, $\om$ and $v$ are all real. All the imaginary parts contained in the mass matrix vanish. Consequently, $R_1$ and $R_3$ mix, but being separated from $I_{1,3}$ and vise versa. We have the physical fields:
\bea h_\eta &=& c_\theta R_1- s_\theta R_3,\hs H_\eta = s_\theta R_1+c_\theta R_3,\\
a_\eta &=& c_\theta I_1 - s_\theta I_3,\hs A_\eta= s_\theta I_1+c_\theta I_3,\eea      
with masses
\bea m^2_{h_\eta}&=& M^2_3,\hs m^2_{a_\eta}=M^2_3,\\ 
m^2_{H_{\eta}}&=&M^2_3+(\la_9+\la_{10})(u^2+\om^2),\hs m^2_{A_{\eta}}=M^2_3+(\la_9-\la_{10})(u^2+\om^2).\eea Here, we have defined $s_\theta\equiv \sin(\theta),\ c_\theta\equiv \cos(\theta)$ and so forth, with
\be t_{\theta}=\fr{u}{\om}. \ee Notice that the $\theta$ is the mixing angle of the charged gauge bosons $W-Y$, which must be small \cite{ecn331}. In the effective limit, we have $h_\eta \simeq R_1$, $a_\eta\simeq I_1$, $H_\eta \simeq R_3$, and $A_\eta\simeq I_3$. The degeneracy of $a_\eta$ and $h_\eta$ masses is due to the fact that a coupling of $\eta$ and $\phi$ similarly to $\la_{10}$ is suppressed by the gauge symmetry, which is unlike the case of the inert doublet model \cite{idmma}. On the other hand, by contrast (which is not the case in the present work) if the $Z_2$ symmetry was spontaneously broken, i.e. $\langle \eta\rangle \neq 0$, the $CP$ would be spontaneously broken too. In such case, the degenerate masses of $a_\eta$ and $h_\eta$ would be separated.      

{\it Normal scalar sector $(\phi,\chi)$:} This section is identical to that of the economical 3-3-1 model, which can be adapted from \cite{ecn331} as given below, for convenience in reading. There are 12 real scalar fields in total for this sector, in which eight of them are Goldstone bosons eliminated by the corresponding eight massive gauge bosons as associated with the broken gauge generators $[SU(3)_L\otimes U(1)_X]/U(1)_Q$. There remain four physical scalar fields, one charged and two neutral, respectively obtained by 
\bea H^\pm&=&s_\xi \chi^\pm_2+c_\xi (s_\theta \phi^\pm_1+c_\theta \phi^\pm_3),\hs m^2_{H^\pm}=\la_7(u^2+v^2+\om^2)\simeq \la_7 \om^2,\\
h&=&c_\zeta S_2 -s_\zeta (s_\theta S_1 +c_\theta S_3),\crn m^2_h&=&2\la_1v^2+2\la_2(u^2+\om^2)-2\sqrt{[\la_1 v^2-\la_2(u^2+\om^2)]^2+\la^2_4v^2(u^2+\om^2)}\crn
&\simeq& \fr{4\la_1\la_2-\la^2_4}{\la_2} v^2,\\
H&=&s_\zeta S_2 +c_\zeta (s_\theta S_1 +c_\theta S_3),\crn m^2_H&=&2\la_1v^2+2\la_2(u^2+\om^2)+2\sqrt{[\la_1 v^2-\la_2(u^2+\om^2)]^2+\la^2_4v^2(u^2+\om^2)}\crn
&\simeq& 4\la_2 \om^2,
\eea   where we have defined \be \phi^0_2=\fr{S_2+iA_2}{\sqrt{2}},\hs \chi^0_1=\fr{S_1+iA_1}{\sqrt{2}},\hs \chi^0_3=\fr{S_3+iA_3}{\sqrt{2}},\ee and 
\be t_{\xi}=\fr{m_W}{m_X}=\fr{v}{\sqrt{u^2+\om^2}}\simeq \fr{v}{\om},\hs t_{2\zeta}=\fr{\la_4 t_\xi}{\la_2-\la_1 t^2_\xi}\simeq (\la_4/\la_2)t_\xi.\ee The mixing angles $\xi$ and $\zeta$ must be small. The $h$ is the standard model like Higgs boson. The $H$ and $H^\pm$ are the new Higgs bosons with respective masses in $\om$ scale. The Goldstone bosons are $G_Z=A_2$, $G_{Z'}=A_3$, $G^{0/0*}_{X}=(G_4\pm iA_1)/\sqrt{2}$ with $G_4=c_\theta S_1-s_\theta S_3$, $G^\pm_W=c_\theta \phi^\pm_1 -s_\theta \phi^\pm_3$, and $G^\pm_Y=c_\xi\chi^\pm_2-s_\xi (s_\theta\phi^\pm_1+c_\theta \phi^\pm_3)$. In the effective limit, we can summarize \cite{ecn331}
\bea \phi &\simeq& \left(\begin{array}{c}
G^+_W\\
v+\fr{1}{\sqrt{2}}(h+i G_Z)\\
H^+\end{array}\right),\hs 
 \chi \simeq \left(\begin{array}{c}
u+G_X\\
G^-_Y\\
\om +\fr{1}{\sqrt{2}}(H+iG_{Z'})\end{array}\right).\eea  

Let us remind that $M^2_3\equiv \mu^2_3+\la_5v^2+\la_6(u^2+\om^2)$ is always in the scale of $\om^2$, independent of if $\mu^2_3$ is in the weak scale $v^2$ or in the 3-3-1 scale $\om^2$. Therefore, all the inert particles in this model are always heavy ($\sim \om$). This is different from the inert doublet model, where the inert particles are naturally in the weak scale. Depending on the relations of $\la_9$, $\la_{10}$ and their signs, we have which inert particle is lightest or LIP. There are three cases:
\ben \item $H_\eta$ is LIP: $\la_{10}<\mathrm{Min}\{0,-\la_9\}$.
\item $A_\eta$ is LIP: $\la_{10}>\mathrm{Max}\{0,\la_9\}$.
\item $h_\eta$ and $a_\eta$ are LIP: $-\la_9 < \la_{10} < \la_9$.    
\een Because $h_\eta$ and $a_\eta$ are degenerate in mass, the third case may be ruled out by the direct detection experiments due to their scattering with nuclei via $Z$ exchange channel \cite{lhall}, which is unlike the inert doublet model. The first and second cases are realistic, which are only existed by the 3-3-1 model. However, in the following we consider only the first case with the dark matter $H_\eta$. For the second case with $A_\eta$, the calculations can be done similarly.  

To close this section, let us calculate the interactions between the two sectors, inert and normal. As mentioned, the inert scalars interact only with normal scalars and gauge bosons, not with fermions. The effect of mixings such as $\theta$, $\zeta$ and $\xi$ will be neglected in the present work since they give very small contributions due to the constraints $u\ll v \ll \om$. The scalar interactions are obtained as follows \bea V_{\mathrm{normal-inert}} &=& \left[(\la_5+\la_8)\left(\sqrt{2}v h +\fr{h^2}{2}\right)+\la_5 H^+ H^- +\la_6 \left(\sqrt{2}\om H +\fr{H^2}{2}\right)\right]h^+_\eta h^-_\eta\crn
&&+\left[\la_5\left(\sqrt{2}vh +\fr{h^2}{2}\right)+\la_5 H^+ H^- +\la_6 \left(\sqrt{2}\om H +\fr{H^2}{2}\right)\right]\fr{a^2_\eta +h^2_\eta}{2}\crn
&&+  \left[\la_5 \left(\sqrt{2}v h +\fr{h^2}{2}\right)+(\la_5+\la_8)H^+ H^- \right.\crn
&&\hs \left.+(\la_6+\la_9+\la_{10})\left(\sqrt{2}\om H +\fr{H^2}{2}\right) \right]\fr{H^2_\eta}{2}\crn
&&+  \left[\la_5 \left(\sqrt{2}v h +\fr{h^2}{2}\right)+(\la_5+\la_8)H^+ H^- \right.\crn
&&\hs \left.+(\la_6+\la_9-\la_{10})\left(\sqrt{2}\om H +\fr{H^2}{2}\right) \right]\fr{A^2_\eta}{2}\crn
&&+\fr{u(\la_9+\la_{10})}{\sqrt{2}}Hh_\eta H_\eta +\fr{u(\la_9-\la_{10})}{\sqrt{2}}Ha_\eta A_\eta \crn
&&+\left[\fr{\la_8}{2}(\sqrt{2}v+h)H^+h^-_\eta (H_\eta-iA_\eta)+H.c.\right]\label{scalar-inert} \eea
The identification of gauge bosons can be found in \cite{ecn331}. Hence, the interactions between the inert scalars and gauge bosons can be derived from the Lagrangian (\ref{latoanphan}). The triple interactions of two inert scalars and one gauge boson are 
\bea \mathcal{L}^{\mathrm{triple}}_{\mathrm{gauge-inert}}&=&\fr{g}{2}\left(\fr{1}{c_W}Z^\mu+\fr{c_{2W}}{c_{W}\sqrt{3-4s^2_W}}Z'^\mu\right)h_\eta \overleftrightarrow{\pa_\mu} a_\eta -g\fr{c_W}{\sqrt{3-4s^2_W}}Z'^\mu H_\eta \overleftrightarrow{\pa_\mu}A_\eta\crn
&&+i\fr{g}{2}\left(-2s_W A^\mu -\fr{c_{2W}}{c_W}Z^\mu +\fr{c_{2W}}{c_W\sqrt{3-4s^2_W}}Z'^\mu\right)h^-_\eta \overleftrightarrow{\pa_\mu} h^+_\eta \crn 
&&+\fr{g}{\sqrt{2}}\left(iW^{+\mu} h^-_\eta \overleftrightarrow{\pa_\mu}\fr{h_\eta-i a_\eta}{\sqrt{2}} +iX^{0\mu}\fr{H_\eta +i A_\eta}{\sqrt{2}}\overleftrightarrow{\pa_\mu}\fr{h_\eta - i a_\eta}{\sqrt{2}}\right.
\crn
&&\left. 
+iY^{-\mu}\fr{H_\eta+iA_\eta}{\sqrt{2}}\overleftrightarrow{\pa_\mu}h^+_\eta +H.c.\right),\label{triple} \eea where we have denoted $A \overleftrightarrow{\pa_\mu}B=A(\pa_\mu B)-(\pa_\mu A)B$. The quartic interactions of two inert scalars and two gauge bosons are given by 
\bea \mathcal{L}^{\mathrm{quartic}}_{\mathrm{gauge-inert}}&=&\fr{g^2}{2}\left[\fr 1 2 \left(\fr{1}{c_W}Z_\mu+\fr{c_{2W}}{c_W\sqrt{3-4s^2_W}}Z'_\mu\right)^2+W^+_\mu W^{-\mu}+X^{0*}_\mu X^{0\mu}\right]\fr{h^2_\eta+a^2_\eta}{2}\crn
&&+\fr{g^2}{2}\left[\fr 1 2 \left(-2s_W A_\mu -\fr{c_{2W}}{c_W}Z_\mu +\fr{c_{2W}}{c_W\sqrt{3-4s^2_W}}Z'_\mu \right)^2 + W^+_\mu W^{-\mu}\right. \crn
&&\left.+ Y^+_\mu Y^{-\mu}\right] h^+_\eta h^-_\eta
+ \fr{g^2}{2}\left[\fr{2c^2_W}{3-4s^2_W}Z'_\mu Z'^\mu+ X^{0*}_\mu X^{0\mu}+Y^+_\mu Y^{-\mu}\right]\fr{H^2_\eta + A^2_\eta}{2}\crn
&&+\fr{g^2}{2\sqrt{2}}\left\{\left[2\left(-s_W A_\mu+\fr{s^2_W}{c_W}Z_\mu +\fr{c_{2W}}{c_W\sqrt{3-4s^2_W}}Z'_\mu\right)W^{+\mu}+\sqrt{2}X^0_\mu Y^{+\mu}\right]\right.\crn
&&\times \fr{h_\eta-i a_\eta}{\sqrt{2}}h^-_\eta + \left[\left(\fr{1}{c_W}Z_\mu-\fr{1}{c_W\sqrt{3-4s^2_W}}Z'_\mu\right)X^{0\mu}+\sqrt{2}W^+_\mu Y^{-\mu}\right]\crn
&&\times \fr{h_\eta - i a_\eta}{\sqrt{2}}\fr{H_\eta + iA_\eta}{\sqrt{2}}+\left[-\left(2s_W A_\mu +\fr{c_{2W}}{c_W}Z_\mu +\fr{1}{c_W\sqrt{3-4s^2_W}}Z'_\mu\right)Y^{-\mu}\right.\crn
&&\left.\left. +\sqrt{2}W^-_\mu X^{0\mu}\right] h^+_\eta \fr{H_\eta+i A_\eta}{\sqrt{2}}+H.c. \right\}\label{quartic}\eea Let us remind the reader that in (\ref{triple}) and (\ref{quartic}), the $A_\mu$ is photon field, the $Z_\mu$ and $W^\pm_\mu$ are standard model like. Whereas, the $Z'_\mu$ is a new neutral gauge boson and the $X^{0,0*}_\mu$, $Y^\pm_\mu$ are new non-Hermitian gauge bosons. From (\ref{scalar-inert}), (\ref{triple}) and (\ref{quartic}), we explicitly see that the inert particles are only coupled in pairs in the interactions, as predicted. Also, the Feynman rules due to these interactions as used below are easily achieved, which should be understood. The ordinary Feynman rules of the economical 3-3-1 model can be found in \cite{ecn331}.

\section{\label{relicdensity} Dark matter constraint}

\subsection{Relic density} 
We can discuss in two cases: (i) $H_\eta$ is lighter than every new particle of the economical 3-3-1 model such as $H$, $H^\pm$, $Z'$, $X$, $Y$, $U$, $D$ and $\nu_R$; (ii) By contrast, $H_\eta$ is heavier than some or all those new particles. Which case presents depending on the parameter space ($\mu_3$, $\om$, $\la_{2,6,7,9,10}$, and $h^{U,D}$) of the model.  In the first case, the contribution to the dark matter relic density includes only the annihilation processes of dark matter into the standard model particles. Whereas, in the second case the dark matter can be annihilated into the new particles of the economical 3-3-1 model, which may dominate over the standard model productions. For our purpose, in this work it is sufficiently to consider only the first case. Also, the coannihilation of $H_\eta$ with any of $A_\eta$, $h_\eta$, $a_\eta$ and $h^\pm_\eta$ will be neglected. The second case needs more study to be published elsewhere.               

\begin{figure}[h]
\begin{center}
\includegraphics{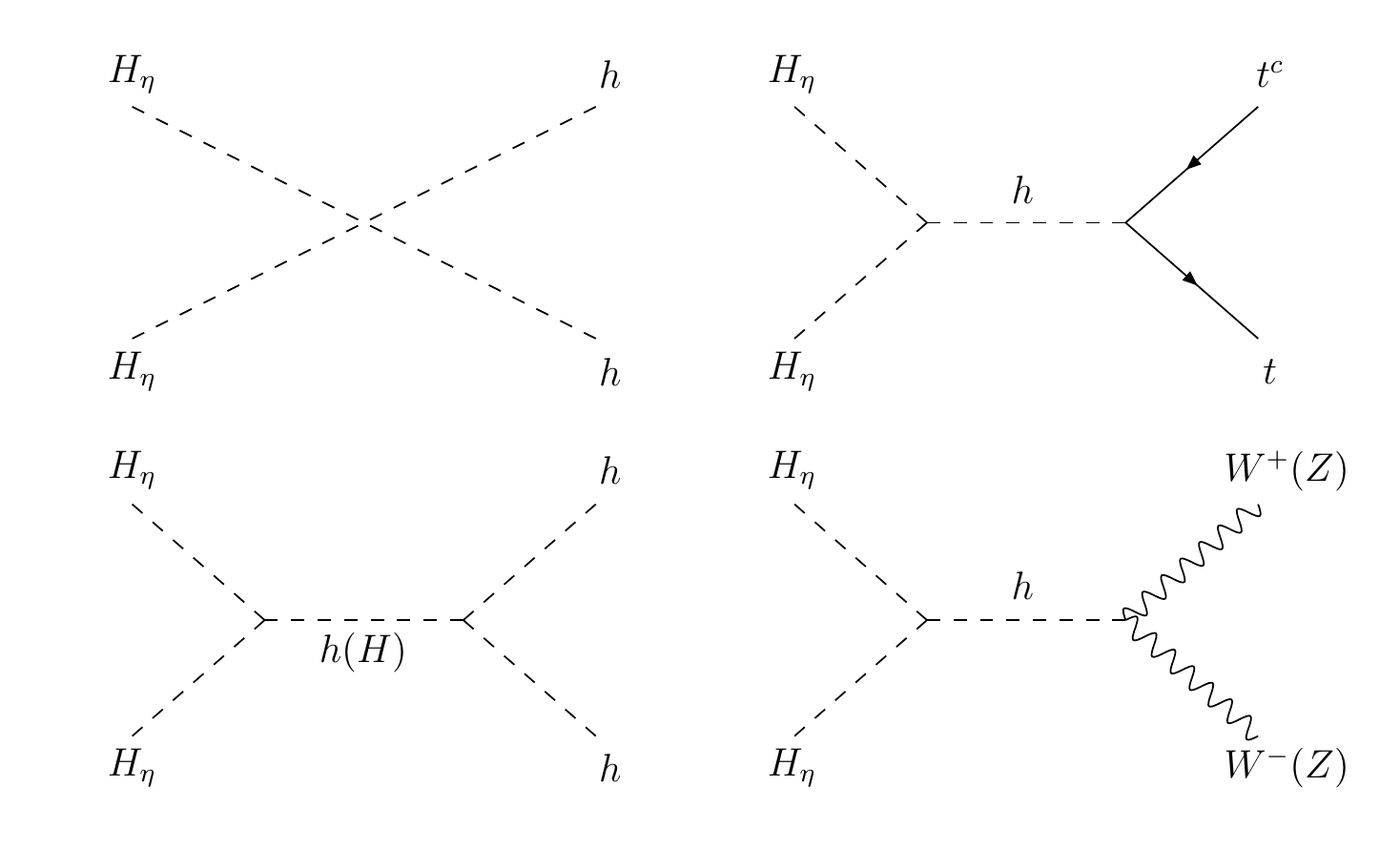}
\caption[]{\label{case1} Dominant contributions to $H_\eta$ annihilation when it is lighter than the new particles of the economical 3-3-1 model.}
\end{center}
\end{figure}

The dominant contibutions to the relic density of dark matter $H_\eta$ come from the diagrams as given in Fig. \ref{case1}.  
The thermal average on the cross-section times relative velocity between two incoming dark matters is obtained by  
\bea \langle \sigma v_{\mathrm{rel}}\rangle &=& \fr{1}{64\pi m^2_{H_\eta}}\left(1-\fr{6}{x_F}-\fr 1 2 \fr{m^2_h}{m^2_{H_\eta}}\right)\left[\la_5-3\la_5\fr{m^2_h}{4m^2_{H_\eta}}\left(1+\fr{m^2_h}{4m^2_{H_\eta}}\right) -2\la_4\fr{m^2_{H_\eta}-\mu^2_3}{4m^2_{H_\eta}-m^2_H}\right]^2\crn
&& + \fr{3}{16\pi x_F m^2_{H_\eta}}\left(1-\fr 1 2 \fr{m^2_h}{m^2_{H_\eta}}\right) \left[\la_5-3\la_5\fr{m^2_h}{4m^2_{H_\eta}}\left(1+\fr{m^2_h}{4m^2_{H_\eta}}\right) -2\la_4\fr{m^2_{H_\eta}-\mu^2_3}{4m^2_{H_\eta}-m^2_H}\right]
\crn && \times \left[3\la_5\fr{m^2_h}{4m^2_{H_\eta}}+8\la_4\fr{(m^2_{H_\eta}-\mu^2_3) m^2_{H_\eta}}{(4m^2_{H_\eta}-m^2_H)^2}\right]
+\fr{\la^2_5 }{64\pi}\left[\fr{m^4_W}{m^6_{H_\eta}}\left(1-\fr{18}{x_F}-\fr 1 2 \fr{m^2_W-m^2_h}{m^2_{H_\eta}}\right)\right.\crn
&&\left.+\fr{2}{m^2_{H_\eta}}\left(1-\fr{6}{x_F}-\fr{3m^2_W-m^2_h}{2m^2_{H_\eta}}\right)\right] +\fr{\la^2_5 }{128\pi}\left[\fr{m^4_Z}{m^6_{H_\eta}}\left(1-\fr{18}{x_F}-\fr 1 2 \fr{m^2_Z-m^2_h}{m^2_{H_\eta}}\right)\right.\crn 
&&\left. +\fr{2}{m^2_{H_\eta}}\left(1-\fr{6}{x_F}-\fr{3m^2_Z-m^2_h}{2m^2_{H_\eta}}\right)\right] +\fr{3\la^2_5 m^2_t}{64\pi m^4_{H_\eta}}\left(1-\fr{12}{x_F}- \fr{3m^2_t}{2m^2_{H_\eta}}+\fr{m^2_h}{2 m^2_{H_\eta}}\right).\label{crsv} \eea Here, we have used the fact that the $H_\eta$ is non-relativistic, and the result was given as an expansion up to the squared velocity of $H_\eta$ with $\langle v^2\rangle=6/x_F$ and $x_F=m_{H_\eta}/T_F\sim 20$ at the freeze-out temperature \cite{reliccal}. Also, we have utilized the approximation: $m^2_{H_\eta}-\mu^2_3 = \la_5 v^2+ (\la_6+\la_9+\la_{10})(u^2+\om^2)  \simeq (\la_6+\la_9+\la_{10})\om^2$ due to $u^2,\ v^2 \ll \om^2$. 

Because $H_\eta$ is lighter than the new particles of the economical 3-3-1 model (with the masses $\sim \om$), it is strongly to impose $\mu^2_3\ll \om^2$, i.e. $(\la_6+\la_9+\la_{10})\om^2 \simeq m^2_{H_\eta}-\mu^2_3\simeq  m^2_{H_\eta}$. Therefore, the parameter space in the first case is given by appropriate conditions on the coupling $\la_6+\la_9+\la_{10}$. For example, for $H$ with mass $m^2_H\simeq 4\la_2 \om^2$, the condition is $\la_6+\la_9+\la_{10} < 4\la_2$. However, it is noticed that the following discussions are unchanged for any size of $\mu^2_3$ that satisfies the present case. Because the dark matter $H_\eta$ is naturally heavy in $\om$ scale, the ratios $\fr{m^2_{W}}{m^2_{H_\eta}}$, $\fr{m^2_{Z}}{m^2_{H_\eta}}$, $\fr{m^2_{h}}{m^2_{H_\eta}}$, $\fr{m^2_{t}}{m^2_{H_\eta}}$ are negligible that  can be terminated in the effective limit. Hence, the result (\ref{crsv}) can be approximated as 
\be \langle \sigma v_{\mathrm{rel}}\rangle \simeq \fr{\al^2}{(150\ \mathrm{GeV})^2}\left(\fr{\la_5\times 1.92\ \mathrm{TeV}}{m_{H_\eta}}\right)^2\left(1.04+0.35a^2+2.39ab\right),\label{crsv1}\ee  where $\al\simeq 1/128$ is the fine structure constant, $x_F=20$ has been used, and 
\be a\equiv 1-2\fr{\la_4}{\la_5}\fr{m^2_{H_\eta}}{4m^2_{H_\eta}-m^2_H},\hs b\equiv \fr{\la_4}{\la_5}\fr{m^4_{H_\eta}}{(4m^2_{H_\eta}-m^2_H)^2}.\label{ab12} \ee 

The dark matter density can be evaluated as $\Om_{H_\eta}h^2\simeq 0.1\ \mathrm{pb}/\langle \sigma v_{\mathrm{rel}}\rangle$ \cite{reliccal}, which depends on only four parameters such as $m^2_{H_\eta}$, $m^2_H$, $\la_5$ and $\la_4$ because of (\ref{crsv}) or alternatively $m_{H_\eta}/\la_5$, $a$ and $b$ due to (\ref{crsv1}). Since $\fr{\al^2}{(150\ \mathrm{GeV})^2}\simeq 1\ \mathrm{pb}$, the WMAP data $\Om_{H_\eta}h^2\simeq 0.11$ \cite{pdg} imply 
\be m_{H_\eta}\simeq \la_5 \times \sqrt{1.04+0.35a^2+2.39ab} \times 2\ \mathrm{TeV}.\label{md1}\ee Because $H$ and $H_\eta$ have the masses in $\om$ scale, the $a$ and $b$ can be naturally given in order of unity [their correct values can be derived from (\ref{ab12}) that depend on only the scalar coupling ratios, $\la_4/\la_5$ and $(\la_6+\la_{9}+\la_{10})/\la_2$]. Moreover, the $\la_5$ coupling is constrained by $0<\la_5<8\pi$ (the right inequality exists if we require the potential to be perturbative) which is in order of unity also. Consequently, the dark matter $H_\eta$ have a right relic density with its mass naturally in TeV scale due to (\ref{md1}), $m_{H_\eta}= \mathcal{O}(1)\ \mathrm{TeV}$. To be concrete, let us give an estimation as follows. Since, in the present case, $H_\eta$ considered is lightest among the new particles including $H$, we can suppose that $m^2_H$ is large enough in comparison to that of $H_\eta$ so that the squared mass ratios in $a$ and $b$ are negligible, thus $a\simeq 1$ and $b\simeq 0$ (this also applies when $H$ does not couple to $h$, i.e. $\la_4=0$). Therefore, we have \be m_{H_\eta}\simeq \la_5\times 2\ \mathrm{TeV},\label{md2}\ee which is around 2 TeV if one takes $\la_5$ about one.                
            
The inert doublet model provides a LIP dark matter possibly in weak either TeV scale. However, our model indicates to the LIP dark matter only in TeV scale, behaving as a scalar singlet under the standard model symmetry. The TeV mass of dark matter in our model is a natural consequence of the 3-3-1 symmetry breaking scale ($\om$). However, in the inert doublet model, since there is only a scale $v$ the large mass is only enhanced by the large scalar coupling, which reaches the applicable limit of perturbative theory. In this case, the normal sector and inert sector become strongly coupled, which contradicts to our case with usual scalar couplings, for example, $\la_5\sim 1$ as clarified above.            

\subsection{Direct search}

\begin{figure}[h]
\begin{center}
\includegraphics{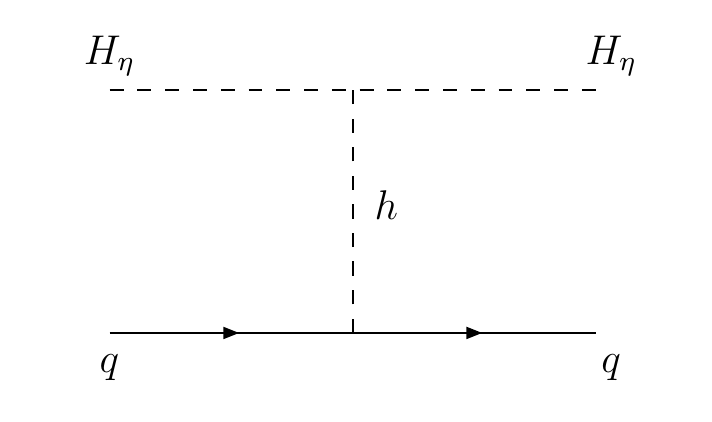}
\caption[]{\label{dddmf} Dominant contributions to $H_\eta$-quark scattering.}
\end{center}
\end{figure}

The direct dark matter search measures the recoil energy deposited by the dark matter scattering on nuclei of a large detector. This scattering is due to the interactions of dark matter with quarks confined in nucleons. Since the dark matter is very non-relativistic, the process can be described by the effective Lagrangian \cite{bbpsdd},
\be \mathcal{L_S}=2\la_q m_{H_\eta} H_\eta H_\eta \bar{q}q.\ee Note that, for the real scalar field only spin-independent and even interactions are possible. The effective interaction above can be obtained by the t-channel exchange of $h$ as the diagram depicted by Fig. \ref{dddmf}. 
Therefore, we have 
\be \la_q=\fr{\la_5 m_q}{2m_{H_\eta}m^2_h}.\label{hsq}\ee 

The $H_\eta$-nucleon scattering amplitude can be given as a summation over the quark level interactions with respective nucleon form factors. The $H_\eta$-nucleon cross-section is      
\be \sigma_{H_\eta-N}=\fr{4m^2_{r}}{\pi}\la^2_N, \ee where $N= p,\ n$ denotes nucleon, and 
\bea m_r &=& \fr{m_{H_\eta}m_N}{m_{H_\eta}+m_N}\simeq m_N,\crn 
\fr{\la_N}{m_N}&=&\sum_{u,d,s}f^N_{Tq}\fr{\la_q}{m_q}+\fr{2}{27}f^N_{TG}\sum_{c,b,t}\fr{\la_q}{m_q}\simeq 0.35\fr{\la_5}{2m_{H_\eta}m^2_h},\eea with $f^N_{TG}=1-\sum_{u,d,s}f^N_{Tq}$ and the $f^N_{Tq}$ values have been taken from \cite{john}. Let $m_N=1$ GeV and $m_h=125$ GeV \cite{atlas,cms}. We have    
\be \sigma_{H_\eta-N} \simeq \left(\fr{\la_5\times 2\ \mathrm{TeV}}{m_{H_{\eta}}}\right)^2\times 1.56 \times 10^{-44}\ \mathrm{cm}^2.\ee Since the value in parenthesis is in order of unity as given above, the cross-section is in good agreement with the XENON100 experimental data \cite{xenon100}. If the mass of $H$ is much larger than $H_\eta$, the model predicts 
\be \sigma_{H_\eta-N} = 1.56 \times 10^{-44}\ \mathrm{cm}^2, \ee for the dark matter with mass in TeV range.

 \section{\label{why} The necessity of this work and its implication}

We have given a discussion on dark matter search status in the 3-3-1 models in \cite{dongdm}. Here we will provide a detailed analysis in order to show explicitly why this work is needed. Its signification for solving dark matter problem in typical 3-3-1 models is also given.  

\subsection{Why this work is needed}

As a result of 3-3-1 gauge symmetry and its particle content, the gauge interactions, minimal Yukawa Lagrangian and minimal scalar potential of the theory normally couple new particles concerned in pairs in interacting with the standard model particles, similarly to superparticles in supersymmetry \cite{dongdm,dm331}. Therefore, the extended sectors in 3-3-1 models such as scalar, fermion and gauge have usually been thought to provide some candidates for dark matter. However, the problem encountered is how to suppress the unwanted interactions and vacuums \cite{dongdm,lepto331}, which lead to the fast decay of dark matter. In the typical 3-3-1 models \cite{331r,331m}, the new particles concerned are bileptons and the unwanted interactions (other than the minimal interactions) are the ones that violate the lepton number \cite{lepto331}. In the 3-3-1 model with right-handed neutrinos, the unwanted vacuums are the ones when neutral scalar bileptons such as $\chi^0_1$ and $\eta^0_3$ develop nonzero VEVs.    

The first three articles of \cite{dm331} were the first works on identifying dark matter candidates in 3-3-1 models. However, their stability mechanism was not given. The first article of \cite{dm331} discussed dark matter in the minimal 3-3-1 model, however it gave a wrong identification of dark matter. In fact, the candidate obtained therein (which is similar to the imaginary part of $\chi^0_3$ in the text) is the Goldstone boson of $Z'$ gauge boson which is an unphysical particle. Even if the corresponding Higgs scalar as mentioned therein (which is similar to the real part of $\chi^0_3$) was interpreted as a dark matter, it will decay into the standard model particles via the tree-level coupling of the candidate to the standard model Higgs bosons $\sim \om \mathrm{Re}(\chi^0_3) h h$ (since it has a VEV $\om$). As a fact, the minimal 3-3-1 model in its current form may contain no dark matter.  

The second and third articles of \cite{dm331} gave a discussion of dark matter in the 3-3-1 model with right-handed neutrinos. The candidates identified were the real and/or imaginary parts of $\eta_3$ as in the text. However, as mentioned what is the mechanism for dark matter stability was not provided. Hence, there is no reason why $\eta^0_3$ (even $\chi^0_1$) cannot develop a VEV as well as the lepton number violating interactions are turned on, which lead to the tree-level couplings of dark matter with the standard model particles. For examples, when $\eta^0_3$ develops a VEV, its real part will decay into two standard model Higgs bosons. Moreover, both real and imaginary parts will decay into light quarks due to the mixing of ordinary and exotic quarks. The presence of lepton number violating Yukawa interactions will lead to the decay of candidate into light quarks due to the same reason of the previous example. Whereas, the lepton number violating scalar potential would lead to the tree-level coupling of candidate to the two standard model Higgs bosons. In addition, the neutral scalar bileptons including the candidate might develop VEVs due to these violating scalar interactions.       

To solve the above problems, the fourth article of Ref. \cite{dm331} was the first one introducing the extra symmetry for dark matter stability in 3-3-1 models. It studied the 3-3-1 model with right-handed neutrinos above and regarding the lepton number symmetry as a mechanism for dark matter stability. It was intriguing that this symmetry would suppress all the unwanted interactions and vacuums, which violate or break the lepton number. There, the lightest bilepton particle (possibly $\eta^0_3$ as assumed in the fourth article of \cite{dm331}) was predicted to be stabilized responsible for dark matter. However, the problem was to generate the mass for neutrinos. As \cite{diaspiressilva} cited therein, the neutrinos would get masses from five-dimensional effective interactions which explicitly violate the lepton number (it was a contradiction to the postulate). In fact, these interactions will lead to the fast decay of dark matter into light neutrinos because there are mixings between right-handed and left-handed neutrinos.   

To overcome the above difficulty, the fifth article of Ref. \cite{dm331} introduced another lepton sector (the model was changed and called as the 3-3-1 model with left-handed neutrinos) so that the bilepton character of the new particles is lost. The lepton number symmetry takes no role in stabilizing dark matter. Instead, a $Z_2$ symmetry or $U(1)_G$ were included. The $Z_2$ must be broken by the Higgs vacuum. Therefore, there is no reason why the dark matter $\eta^0_3$ that carries no lepton number cannot develop a VEV and decay then. On the other hand, the $U(1)_G$ must be broken due to its nontrivial dynamics as shown in \cite{dongdm}. It cannot prevent the dark matter from the decay. A suggestion in \cite{dongdm} was that $G$-parity, $(-1)^G$, may be a mechanism for dark matter stability. In Ref. \cite{dongdm}, we gave a mechanism for dark matter stability based on $W$-parity, similarly to $R$-parity in supersymmetry. However, the dark matter model works only with the fermion content of the 3-3-1 model with neutral fermions. 

To conclude, the problems on dark matter identification and stability in the typical 3-3-1 models, the 3-3-1 model with right-handed neutrinos and the minimal 3-3-1 model, remain unsolved, which have called for our attention.  

Via our work given above, we see that the typical 3-3-1 models are only self-consistent if they contain interactions explicitly violating the lepton number. If one scalar triplet of the 3-3-1 model with right-handed neutrinos is inert ($Z_2$ odd), the other two scalar triplets will result an economical 3-3-1 model self-consistent. This model provides consistent masses for neutrinos \cite{ecn331}. [The neutrinos can get masses via two ways similarly to the economical 3-3-1 model: radiative corrections (as given in the sixth article of \cite{ecn331}), or alternatively effective interactions (as given in the sixth and eighth article of \cite{ecn331}). In all these ways, the neutrino masses are generated due to the contribution of only $\chi$ and $\phi$ ($Z_2$ even scalars), while the $\eta$ does not contribute due to $\langle \eta \rangle = 0$ under the $Z_2$ symmetry. The generation of neutrino masses is also accompanied with the interaction of inert scalars ($\eta$) with leptons ($\psi$).  
But, since the theory conserves the $Z_2$ symmetry, the inert scalars ($\eta$) which are odd under $Z_2$ are only coupled in pairs in such interactions. For example, an effective interaction can be included as $\bar{\psi}^c_L \psi_L (\eta\eta)^*$ and its hermitian conjugation, which leads to the interactions, $\psi_i \psi_j \eta^*_i \eta^*_j$, where $i,j$ are $SU(3)_L$ indices. Since all the components $\eta_i$ are odd under the $Z_2$, this may lead to the decay of an inert scalar ($\eta_i$) with larger mass into another inert scalar ($\eta_j$) with smaller mass (associated with two leptons $\psi_i\psi_j$). In other words, the transitions or decays ($\eta_i\leftrightarrow\eta_j$) happen only in the dark sector of inert particles. The LIP ($H_\eta$) cannot decay into other inert particles (which have larger masses) due to kinematically suppressed as well as cannot decay into the normal particles of the economical 3-3-1 model due to the $Z_2$ symmetry. It is absolutely stabilized. Let us remind the reader that in the model of fourth article of \cite{dm331}, such similar interactions happen, by contrast, between the $\eta_3$ of the assumed dark sector (the bilepton particles) and usual particles $\eta_{1,2}$ (which carry no lepton number and couple to the standard model particles, even $\eta_1$ develops the VEV) of the normal matter sector, which subsequently lead to the fast decay of $\eta_3$. The candidate is unstable.] The dark matter thus results as resided in the inert part of the model as given above. Although our candidates $H_\eta$ and $A_\eta$ are similar to those ($\eta^0_3$) as studied in previous literature \cite{dm331,dongdm}, its phenomenology is completely distinguished. This is due to
\ben
\item The masses of $H_\eta$ and $A_\eta$ are separated due to the lepton number violating coupling $\la_{10}$. They are two distinct particles. In the previous studies, their masses are degenerate \cite{dm331,dongdm}. In fact, they are different components of a complex field $\eta^0_3$. 
\item $H_\eta$ and $A_\eta$ do not couple to fermions. However, those in \cite{dm331,dongdm} do. 
\item $H_\eta$ and $A_\eta$ work in the economical 3-3-1 model with lepton number violations and the neutrino masses got naturally generated \cite{ecn331}. Those in \cite{dm331,dongdm} work in different 3-3-1 models. In addition, for the 3-3-1 model with right-handed neutrinos we cannot understand the physics of assumed dark matter $\eta^0_3$ and neutrino masses simultaneously. The model is in fact unrealistic as indicated above.        
\een     
Finally, we can have other cases of inert scalar triplet as given below. In these cases, the dark matter candidates completely differ from $\eta^0_3$.

\subsection{Implication of this work}

For the 3-3-1 model with right-handed neutrinos, we can introduce another scalar sector which can provide dark matter. That is, 
$\phi$ and $\chi$ are the same model proposed above, however the inert triplet is changed to $\eta=(\eta^+_1,\ \eta^0_2,\ \eta^+_3)\sim (1,3,2/3)$ which is a replication of $\phi$. In this case, we can have a doublet dark matter similarly to the inert doublet model. 

For the minimal 3-3-1 model, the scalar sector is $\rho=(\rho^+_1,\ \rho^0_2,\ \rho^{++}_3)\sim (1,3,1)$, $\eta=(\eta^0_1,\ \eta^-_2,\ \eta^+_3)\sim (1,3,0)$, and $\chi=(\chi^-_1,\ \chi^{--}_2,\ \chi^0_3)\sim (1,3,-1)$. The reduced 3-3-1 model works with only $\rho$ and $\chi$ by removing $\eta$ either works with $\eta$ and $\chi$ by removing $\rho$ \cite{rm331}. Therefore, we have the following cases for dark matter of the minimal 3-3-1 model: 
\ben
\item $\eta$ is inert scalar triplet. We may have a doublet dark matter similarly to the inert doublet model.
\item $\rho$ is inert scalar triplet. A doublet dark matter may result, similarly to the previous case.
\item Removing $\eta$ ($\rho$), we introduce the inert scalar triplet as a replication of $\rho$ ($\eta$) instead.
\item Removing $\eta$ or $\rho$, we include the inert triplet as a replication of $\chi$ instead. These cases will yield a singlet dark matter. 
\een       

All the cases above are worth exploring \cite{dongsoa}. Therefore, as an example, in the present work we have presented only one case of the 3-3-1 model with right-handed neutrinos as given above in Sec. \ref{model} and \ref{relicdensity}. 

To summarize, the mechanism given in this work responsible for dark matter stability is a solution to the dark matter problem of the typical 3-3-1 models, the 3-3-1 model with right-handed neutrinos and the minimal 3-3-1 model. The dark matter candidates obtained and their phenomenologies are rich and unlike those in the previous studies \cite{dm331,dongdm}. The resulting 3-3-1 models under this mechanism are self-consistent and the neutrinos get desirable masses.

\section{\label{conclusion}Conclusion}

As a nature of the typical 3-3-1 models, the lepton number appears to be a residual charge that is not commuted with the gauge symmetry. If the lepton number is conserved, it will behave as a local charge. And, the 3-3-1 gauge symmetry should be extended. One way to keep the 3-3-1 models self-consistent (which avoids an extension) is that the lepton number should be belong to an approximate symmetry, and the 3-3-1 models must contain interactions that explicitly violate the lepton number. Looking into the other variants of the 3-3-1 models, we observe that the economical 3-3-1 model is a natural recognition of the above criteria, while the reduced 3-3-1 model \cite{rm331} at renormalizable level is not. However, the reduced 3-3-1 model will be viable when the effective interactions responsible for fermion masses are included.   

We have proved that the 3-3-1 model with right-handed neutrinos can by itself contain an inert scalar triplet ($\eta$) responsible for dark matter, while its remaining part with other multiplets work as in the economical 3-3-1 model. Formerly, the $\eta$ triplet was neglected when one considers the economical 3-3-1 model \cite{ecn331}. The stability of dark matter candidate ($H_\eta$) as contained in $\eta$ is ensured by a $Z_2$ symmetry (assigned so that only $\eta$ is odd; all other multiplets are even) which has been shown to be not broken by the vacuum. Contradicting to the inert doublet model, our dark matter candidate behaves as a singlet under the standard model symmetry. And, this particle is naturally heavy in the $\om$ scale of 3-3-1 symmetry breaking. The interaction between the inert particles as resided in $\eta$ and the economical 3-3-1 model particles have been also given at the effective limit.      

We have calculated the relic density of dark matter for the case that this particle is lightest among the new particles. The relic density will get the correct value in comparison to WMAP data provided that our dark matter candidate is in TeV range as expected for the new physics of 3-3-1 models. In such range of dark matter mass, the dark matter-nucleon scattering cross-section gets also safe values in the bound of the strongest experimental data such as that of XENON100. If the new neutral scalar mass ($H$) is more larger than the dark matter mass, i.e. $m^2_{H_\eta}/m^2_{H}$ is negligible, our model predicts the dark matter mass $m_{H_\eta}=\la_5\times 2$ TeV and the nucleon scattering cross-section $\sigma_{H_\eta-N}=1.56\times 10^{-44}$ cm$^2$, remarkably coinciding with the current bound of direct detection experiments such as XENON100 in the TeV range. 

If the dark matter is heavier than some new particles of the economical 3-3-1 model, it will also annihilate into those new particles for the thermal process, which can dominate. Also, the co-annihilation phenomenology of dark matter with other inert particles is interesting. In addition, the inert scalar triplet can be a replication of $\phi$ instead of the current one, which results a doublet dark matter. All these are devoted to further studies. It is well-known that the minimal 3-3-1 model in its current form does not contain any dark matter candidate. By our proposal, the model can similarly be modified to work as a reduced 3-3-1 model \cite{rm331} while containing an inert scalar triplet responsible for dark matter. The dark matter candidate in such model is a scalar doublet under the standard model symmetry similarly to the inert doublet model either a scalar singlet similarly to our model given in the text. However, its phenomenology is very distinguished \cite{dongsoa}.

Finally, our work is a solution to the long-standing problem of dark matter in the typical 3-3-1 models, the 3-3-1 model with right-handed neutrinos and the minimal 3-3-1 model.                        

\section*{Acknowledgments}
This work is funded by Vietnam National Foundation for
Science and Technology Development (NAFOSTED) under grant number 103.03-2012.80.

\appendix

\end{document}